\documentclass[epj]{svjour}

\usepackage{graphics}

\usepackage{latexsym}
\usepackage{amsmath}
\usepackage{amsfonts}
\usepackage{amscd}

\def\bq{\begin{eqnarray}}
\def\eq{\end{eqnarray}}
\def\eps{\varepsilon}

\begin{document}
\title{Hopf algebra structures in particle physics}
\author{Stefan Weinzierl
\thanks{Heisenberg fellow of the Deutsche Forschungsgemeinschaft}
}
\institute{Max-Planck-Institut f\"ur Physik (Werner-Heisenberg-Institut), F\"ohringer Ring 6, D-80805 M\"unchen, Germany}
\date{Received: date / Revised version: date}
%
\abstract{
In the recent years, Hopf algebras have been introduced to
describe certain combinatorial properties of quantum field
theories.
I will give a basic introduction to these algebras and
review some occurrences in particle physics.
} 
\maketitle
%

\section{Introduction}
\label{sect:intro}

Hopf algebras were introduced in mathematics in 1941 to describe in an unified manner
similar aspects of groups and algebras \cite{Hopf}.
An article by Woronowicz in 1987 \cite{Woronowicz}, which provided explicit examples of non-trivial
Hopf algebras, triggered the interest of the physics community.
In turn, Hopf algebras have been used for integrable systems and quantum groups.
In 1998 Kreimer and Connes re-examined renormalization of quantum field theories and showed
that it can be described by a Hopf algebra structure \cite{Kreimer:1998dp,Connes:1998qv}.
In this talk I review Hopf algebras and its relations to perturbative quantum field theories. 
Application of Hopf algebras to quantum groups or non-commutative field theories are not
covered here.

\section{Hopf algebras}
\label{sect:hopf}

In this section I recall the definition of a Hopf algebra and I discuss several examples.

\subsection{Definition}
\label{subsect:def}

Let $R$ be a commutative ring with unit $1$.
An algebra over the ring $R$ is a $R$-module together with 
a multiplication $\cdot$ and a unit $e$.
We will always assume that the multiplication is associative.
In physics, the ring $R$ will almost always be a field $K$
(examples are the rational numbers ${\mathbb Q}$, the real numbers ${\mathbb R}$ or the complex
number ${\mathbb C}$). In this case the $R$-module will actually be a $K$-vector space.
Note that the unit can be viewed as a map from $R$ to $A$ and that the multiplication
can be viewed as a map from the tensor product $A \otimes A$ to $A$ (e.g. one takes two elements
from $A$, multiplies them and gets one element out). 

A coalgebra has instead of multiplication and unit the dual structures:
a comultiplication $\Delta$ and a counit $\bar{e}$.
The counit is a map from $A$ to $R$, whereas comultiplication is a map from $A$ to
$A \otimes A$.
Note that comultiplication and counit go in the reverse direction compared to multiplication
and unit.
We will always assume that the comultiplication is coassociative.
The general form of the coproduct is
\bq
\Delta(a) & = & \sum\limits_i a_i^{(1)} \otimes a_i^{(2)},
\eq
where $a_i^{(1)}$ denotes an element of $A$ appearing in the first slot of $A \otimes A$ and
$a_i^{(2)}$ correspondingly denotes an element of $A$ appearing in the second slot.
Sweedler's notation \cite{Sweedler} consists in dropping the dummy index $i$ and the summation symbol:
\bq
\Delta(a) & = & 
a^{(1)} \otimes a^{(2)}
\eq 
The sum is implicitly understood. This is similar to Einstein's summation convention, except
that the dummy summation index $i$ is also dropped. The superscripts ${}^{(1)}$ and ${}^{(2)}$ 
indicate that a sum is involved.

A bialgebra is an algebra and a coalgebra at the same time,
such that the two structures are compatible with each other.
Using Sweedler's notation,
the compatibility between the multiplication and comultiplication is express\-ed as
\bq
\label{bialg}
 \Delta\left( a \cdot b \right)
 & = &
\left( a^{(1)} \cdot b^{(1)} \right)
 \otimes \left( a^{(2)} \cdot b^{(2)} \right).
\eq

A Hopf algebra is a bialgebra with an additional map from $A$ to $A$, called the 
antipode ${\cal S}$, which fulfills
\bq
a^{(1)} \cdot {\cal S}\left( a^{(2)} \right)
=
{\cal S}\left(a^{(1)}\right) \cdot a^{(2)} 
= 0 & &\;\;\; \mbox{for} \; a \neq e.
\eq

\subsection{Examples}
\label{subsect:examples}

\subsubsection{The group algebra}
\label{subsubsect:ex0}

Let $G$ be a group and denote by $KG$ the vector space with basis $G$.
$KG$ is an algebra with the multiplication given by the group multiplication.
The counit $\bar{e}$ is given by:
\bq
\bar{e}\left( g\right) 
& = & 1.
\eq
The coproduct $\Delta$ is given by:
\bq
\Delta\left( g\right) 
& = & g \otimes g.
\eq
The antipode ${\cal S}$ is given by:
\bq
{\cal S}\left( g \right) & = & g^{-1}.
\eq
$KG$ is a cocommutative Hopf algebra. 
$KG$ is commutative if $G$ is commutative.

\subsubsection{Lie algebras}
\label{subsubsect:ex1}

A Lie algebra ${\mathfrak g}$ is not necessarily associative nor does it have a unit.
To overcome this obstacle one considers the 
universal enveloping algebra $U({\mathfrak g})$,
obtained from the tensor algebra $T({\mathfrak g})$
by factoring out the ideal
\bq
X \otimes Y - Y \otimes X - \left[ X, Y \right],
\eq
with $X, Y \in {\mathfrak g}$.
The counit $\bar{e}$ is given by:
\bq
\bar{e}\left( e\right) = 1,
& &
\bar{e}\left( X\right) = 0.
\eq
The coproduct $\Delta$ is given by:
\bq
\Delta(e) = e \otimes e, 
& &
\Delta(X) = X \otimes e + e \otimes X.
\eq
The antipode ${\cal S}$ is given by:
\bq
{\cal S}(e) = e, 
& &
{\cal S}(X) = -X.
\eq

\subsubsection{Quantum SU(2)}
\label{subsubsect:ex2}

The Lie algebra $su(2)$ is generated by three generators $H$, $X_\pm$ with
\bq
\left[ H, X_\pm \right] = \pm 2 X_\pm,
&  & 
\left[ X_+, X_- \right] = H. 
\eq
To obtain the deformed algebra $U_q(su(2))$, the last relation is replaced with
\bq
\left[ X_+, X_- \right] & = & \frac{q^H - q^{-H}}{q-q^{-1}}.
\eq
The undeformed Lie algebra $su(2)$ is recovered in the limit $q \rightarrow 1$.
The counit $\bar{e}$ is given by:
\bq
\bar{e}\left( e\right) = 1,
& &
\bar{e}\left( H\right) = \bar{e}\left( X_\pm\right) = 0.
\eq
The coproduct $\Delta$ is given by:
\bq
\Delta(H) & = & H \otimes e + e \otimes H, \nonumber \\
\Delta(X_\pm) & = & X_\pm \otimes q^{H/2} + q^{-H/2} \otimes X_\pm.
\eq
The antipode ${\cal S}$ is given by:
\bq
{\cal S}(H) = -H, 
& &
{\cal S}(X_\pm) = - q^{\pm 1} X_\pm .
\eq

\subsubsection{Symmetric algebras}
\label{subsubsect:ex2a}

Let $V$ be a finite dimensional vector space with basis $\{v_i\}$.
The symmetric algebra $S(V)$ is the direct sum
\bq
S(V) & = & \bigoplus\limits_{n=0}^\infty S^n(V),
\eq
where $S^n(V)$ is spanned by elements of the form $v_{i_1} v_{i_2} ... v_{i_n}$ with
$i_1 \le i_2 \le ... \le i_n$.
The multiplication is defined by
\bq
\left( v_{i_1} v_{i_2} ... v_{i_m} \right) \cdot \left( v_{i_{m+1}} v_{i_{m+2}} ... v_{i_{m+n}} \right)
 & = &
 v_{i_{\sigma(1)}} v_{i_{\sigma(2)}} ... v_{i_{\sigma(m+n)}},
 \nonumber
\eq
where $\sigma$ is the permutation on $m+n$ elements such that 
$i_{\sigma(1)} \le i_{\sigma(2)} \le ... \le i_{\sigma(m+n)}$.
The counit $\bar{e}$ is given by:
\bq
\bar{e}\left( e\right) = 1, \;\;\;
& &
\bar{e}\left( v_1 v_2 ... v_n\right) = 0.
\eq
The coproduct $\Delta$ is given for the basis elements $v_i$ by:
\bq
\Delta(v_i) & = & v_i \otimes e + e \otimes v_i.
\eq
Using (\ref{bialg})  one obtains for a general element of $S(V)$
\bq
\lefteqn{
\Delta\left( v_1 v_2 ... v_n \right)
 =  
  v_1 v_2 ... v_n \otimes e
+ e \otimes v_1 v_2 ... v_n
} \nonumber \\
 & &
+ \sum\limits_{j=1}^{n-1} \sum\limits_\sigma
  v_{\sigma(1)} ... v_{\sigma(j)}
   \otimes 
  v_{\sigma(j+1)} ... v_{\sigma(n)},
\eq
where $\sigma$ runs over all $(j,n-j)$-shuffles. A $(j,n-j)$-shuffle
is a permutation $\sigma$ of $(1,...,n)$ such that
\bq
\sigma(1) < \sigma(2) < ... < \sigma(j)
& \mbox{and} &
\sigma(k+1) < ... < \sigma(n). \nonumber 
\eq
The antipode ${\cal S}$ is given by:
\bq
{\cal S}( v_{i_1} v_{i_2} ... v_{i_n}) & = & (-1)^n v_{i_1} v_{i_2} ... v_{i_n}.
\eq

\subsubsection{Shuffle algebras}
\label{subsubsect:ex3}

Consider a set of letters $A$. A word is an ordered sequence of letters:
\bq
 w & = & l_1 l_2 ... l_k.
\eq
The word of length zero is denoted by $e$.
A shuffle algebra ${\cal A}$ on the vector space of words is defined by
\bq
\left( l_1 l_2 ... l_k \right) \cdot 
 \left( l_{k+1} ... l_r \right) & = &
 \sum\limits_{\mbox{\tiny shuffles} \; \sigma} l_{\sigma(1)} l_{\sigma(2)} ... l_{\sigma(r)},
\eq
where the sum runs over all permutations $\sigma$, which preserve the relative order
of $1,2,...,k$ and of $k+1,...,r$.
The counit $\bar{e}$ is given by:
\bq
\bar{e}\left( e\right) = 1, \;\;\;
& &
\bar{e}\left( l_1 l_2 ... l_n\right) = 0.
\eq
The coproduct $\Delta$ is given by:
\bq
\Delta\left( l_1 l_2 ... l_k \right) 
& = & \sum\limits_{j=0}^k \left( l_{j+1} ... l_k \right) \otimes \left( l_1 ... l_j \right).
\eq
The antipode ${\cal S}$ is given by:
\bq
{\cal S}\left( l_1 l_2 ... l_k \right) & = & (-1)^k \; l_k l_{k-1} ... l_2 l_1.
\eq

\subsubsection{Rooted trees}
\label{subsubsect:ex4}

\begin{figure}
\resizebox{!}{3cm}{
  \includegraphics[150pt,590pt][420pt,700pt]{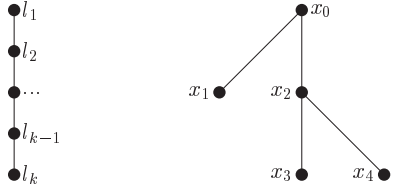}
}
\caption{An element of the shuffle algebra can be represented by a rooted tree without 
side-branchings, as shown in the left figure. 
The right figure shows a general rooted tree with side-branchings. The root is drawn at the top.}
\label{fig:0}
\end{figure}
Consider a set of rooted trees (fig. \ref{fig:0}).
An admissible cut of a rooted tree is any assignment of cuts such that any path from any vertex of the tree 
to the root has at most one cut.
An admissible cut maps a tree $t$ to a monomial in trees $t_1 \times ... \times t_{n+1}$. 
Precisely one of 
these subtrees $t_j$
will contain the root of $t$. We denote this distinguished tree by $R^C(t)$, and the monomial delivered by the $n$ other factors
by $P^C(t)$. 
The counit $\bar{e}$ is given by:
\bq
\bar{e}(e) = 1, \;\;\;
 & &
\bar{e}(t) = 0 \;\;\mbox{for}\; t \neq e.
\eq
The coproduct $\Delta$ is given by:
\bq
\Delta(e) & = & e \otimes e, \\
\Delta(t) & = & t \otimes e + e \otimes t + \sum\limits_{\mbox{\tiny adm. cuts $C$ of $t$}} P^C(t) \otimes R^C(t). \nonumber 
\eq
The antipode ${\cal S}$ is given by:
\bq
{\cal S}(e) & = & e, \nonumber \\
{\cal S}(t) & = & -t - \sum\limits_{\mbox{\tiny adm. cuts $C$ of $t$}} {\cal S}\left( P^C(t) \right) \times R^C(t).
\eq

\subsection{Commutativity and cocommutativity}
\label{subsect:cocomm}

One can classify the examples discussed above into four groups according to whether they are
commutative or cocommutative.
\begin{itemize}
\item Commutative and cocommutative:
Group algebra of a commutative group, symmetric algebras.
\item Non-commutative and cocommutative:
Group algebra of a non-commutative group, universal enveloping algebra of a Lie algebra.
\item Commutative and non-cocommutative:
Shuffle algebra, algebra of rooted trees.
\item Non-commutative and non-cocommutative:
q-deformed algebras.
\end{itemize}
Whereas research on quantum groups focussed primarily on  
non-commutative and non-cocommutative Hopf algebras, it turns out that
for applications in perturbative quantum field theories
commutative, but not necessarily
cocommutative Hopf algebras 
like shuffle algebras, symmetric algebras and rooted trees are the most important ones.

\section{Occurrence in particle physics}
\label{sect:physics}

I will discuss three application of Hopf algebras in perturbative particle physics:
Renormalization, Wick's theorem and Feynman loop integrals.

\subsection{Renormalization}
\label{subsect:renorm}

\begin{figure}
\resizebox{0.5\textwidth}{!}{
  \includegraphics[80pt,500pt][520pt,700pt]{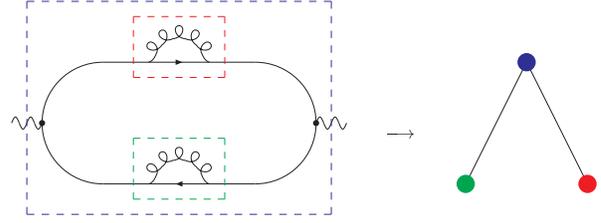}
}
\caption{Nested singularities are encoded in rooted trees.}
\label{fig:1}
\end{figure}

\begin{figure}
\resizebox{0.5\textwidth}{!}{
  \includegraphics[80pt,500pt][595pt,680pt]{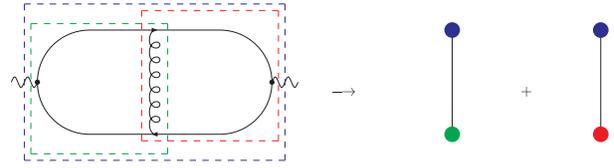}
}
\caption{Overlapping singularities yield a sum of rooted trees.}
\label{fig:2}
\end{figure}

Short-distance singularities of the perturbative expansion of quantum field
theories require renormalization 
\cite{Zimmermann:1969jj}.
The combinatorics involved in the renormalization is governed by a Hopf algebra
\cite{Kreimer:1998dp,Connes:1998qv}
The model for this Hopf algebra is the Hopf algebra of rooted trees (fig. \ref{fig:1} and \ref{fig:2}).


Recall the recursive definition of the antipode:
\bq
{\cal S}(t) & = & -t - \sum\limits_{\mbox{\tiny adm. cuts $C$ of $t$}} {\cal S}\left( P^C(t) \right) \times R^C(t).
\eq
The antipode satisfies
\bq
\label{untwisted}
m \left[ \left( {\cal S} \otimes \mbox{id} \right) \Delta(t) \right] & = & 0,
\eq
where $m$ denotes multiplication:
\bq
m\left( a \otimes b \right) & = & a \cdot b.
\eq
Let 
${\cal R}$ 
be an operation which 
approximates a tree by another tree 
with the same singularity structure
and which satisfies the Rota-Baxter relation:
\bq
\label{rotabaxter}
{\cal R}\left( t_1 t_2 \right) + {\cal R}\left( t_1 \right) {\cal R}\left( t_2 \right) 
 & = &
 {\cal R}\left( t_1 {\cal R}\left( t_2 \right) \right)
 + {\cal R}\left( {\cal R}\left( t_1 \right) t_2 \right).
\nonumber 
\eq
For example, minimal subtraction ($\overline{MS}$) 
\bq
{\cal R}\left( \sum\limits_{k=-L}^\infty c_k \eps^k \right) & = & 
\sum\limits_{k=-L}^{-1} c_k \eps^k
\eq
fulfills the Rota-Baxter relation.
To simplify the notation, I drop the distinction between a Feynman graph
and the evaluation of the graph.
One can now
twist the antipode with ${\cal R}$ and define a new map
\bq
{\cal S}_{\cal R}(t) & = & - {\cal R} \left( 
  t + \sum\limits_{\mbox{\tiny adm. cuts $C$ of $t$}} 
        {\cal S}_{\cal R} \left( P^C(t) \right) \times R^C(t) \right). 
\nonumber
\eq
From the multiplicativity constraint (\ref{rotabaxter}) it follows that
\bq
{\cal S}_{\cal R}\left(t_1 t_2 \right)
 & = &
 {\cal S}_{\cal R}\left(t_1 \right)
 {\cal S}_{\cal R}\left(t_2 \right).
\eq
If we replace ${\cal S}$ by ${\cal S}_{\cal R}$ in (\ref{untwisted})
we obtain
\bq
\label{twisted}
m \left[ \left( {\cal S}_{\cal R} \otimes \mbox{id} \right) \Delta(t) \right] 
 & = & \mbox{finite},
\eq
since by definition ${\cal S}_{\cal R}$ differs from ${\cal S}$ only by
finite terms.
Eq. (\ref{twisted}) is equivalent to the forest formula.
It should be noted that
${\cal R}$ is not unique and 
different choices for ${\cal R}$ correspond
to different renormalization prescription.

\subsection{Wick's theorem}
\label{subsect:wick}

I will discuss here the simplest version of Wick's theorem, which relates
the time-ordered product of $n$ bosonic field operators to the 
normal product of these operators and contractions.
As an example one has
\bq
\lefteqn{
T\left(\phi_1 \phi_2 \phi_3 \phi_4 \right)
 =  
{:\phi_1 \phi_2 \phi_3 \phi_4:}
+ \left( \phi_1, \phi_2 \right) {:\phi_3 \phi_4:}
} \nonumber \\
& &
+ \left( \phi_1, \phi_3 \right) {:\phi_2 \phi_4:}
+ \left( \phi_1, \phi_4 \right) {:\phi_2 \phi_3:}
+ \left( \phi_2, \phi_3 \right) {:\phi_1 \phi_4:}
 \nonumber \\
 & &
+ \left( \phi_2, \phi_4 \right) {:\phi_1 \phi_3:}
+ \left( \phi_3, \phi_4 \right) {:\phi_1 \phi_2:}
+ \left( \phi_1, \phi_2 \right) \left(\phi_3, \phi_4 \right)
 \nonumber \\
 & &
+ \left( \phi_1, \phi_3 \right) \left(\phi_2, \phi_4 \right)
+ \left( \phi_1, \phi_4 \right) \left(\phi_2, \phi_3 \right),
\eq
where I used the notation
\bq
\label{wick1}
\left( \phi_i, \phi_j \right)
 & = & 
 \left\langle 0 \left| T \left( \phi_i \phi_j \right) \right| 0 \right\rangle
\eq
to denote the contraction.
One can use Wick's theorem to define the time-ordered product in terms of the normal
product and the contraction.
To establish the connection with Hopf algebras,
let $V$ be the vector space with basis $\{\phi_i\}$ and identify the normal product
with the symmetric product discussed in sect. \ref{subsubsect:ex2a}
\cite{Fauser:2000yb,Brouder:2002nu}.
This yields the symmetric algebra $S(V)$.
The contraction defines a bilinear form $V \otimes V \rightarrow {\mathbb C}$.
One extends this pairing to $S(V)$ by
\bq
\label{wick2}
\left( {:N_1 N_2:}, M_1 \right)
 & = & \left( N_1, M_1^{(1)} \right) \left( N_2,M_1^{(2)} \right),
 \nonumber \\
\left( N_1, {:M_1 M_2:} \right) 
 & = & \left( N_1^{(1)}, M_1 \right) \left( N_1^{(2)}, M_2 \right).
\eq
Here, $N_1$, $N_2$, $M_1$ and $M_2$ are arbitrary normal products of the $\phi_i$.
With the help of this pairing one defines a new product, called the circle product,
as follows:
\bq
\label{wick3}
N \circ M & = &
 \left( N^{(1)}, M^{(1)} \right) \; {:N^{(2)} M^{(2)}:}
\eq
Again, $N$ and $M$ are normal products.
\begin{figure}
\resizebox{!}{3cm}{
  \includegraphics[50pt,550pt][390pt,710pt]{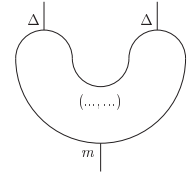}
}
\caption{The ``sausage tangle'': pictorial representation of the definition of the circle product.}
\label{fig:3}
\end{figure}
Fig. \ref{fig:3} shows pictorially the definition of the circle product 
involving the coproduct, the pairing $(...,...)$ and the multiplication.
It can be shown that the circle product is associative.
Furthermore, one obtains that
the circle product coincides with the time-ordered product.
For example,
\bq
\label{wick4}
\phi_1 \circ \phi_2 \circ \phi_3 \circ \phi_4 
 & = &
T\left(\phi_1 \phi_2 \phi_3 \phi_4 \right).
\eq
The reader is invited to verify the l.h.s of (\ref{wick4})
with the help of the definitions (\ref{wick1}), (\ref{wick2}) and (\ref{wick3}). 

\subsection{Loop integrals}
\label{subsect:loop}

\begin{figure}
\resizebox{!}{3cm}{
  \includegraphics[50pt,660pt][260pt,740pt]{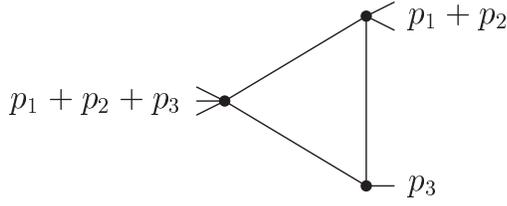}
}
\caption{A one-loop three-point function with two external masses.}
\label{fig:4}
\end{figure}

The calculation of Feynman loop integrals is crucial for 
precise predictions of cross sections in particle physics
phenomenology.
An example is the one-loop three-point function
shown in fig. \ref{fig:4}.
This integral evaluates in dimensional regularization ($D=4-2\eps$) to
a hyper-geometric function:
\bq
\lefteqn{
I(\nu_1,\nu_2,\nu_3) = }
\nonumber \\
& & 
 c_\Gamma
 \left( - s_{123} \right)^{\nu_{123}-D/2}
 \int \frac{d^Dk_1}{i \pi^{D/2}}
 \frac{1}{(-k_1^2)^{\nu_1}}
 \frac{1}{(-k_2^2)^{\nu_2}}
 \frac{1}{(-k_3^2)^{\nu_3}}
 \nonumber \\
 & = &
 c_\Gamma
 \frac{1}{\Gamma(\nu_1)\Gamma(\nu_2)}
 \frac{\Gamma(D/2-\nu_1)\Gamma(D/2-\nu_{23})}{\Gamma(D-\nu_{123})}
 \nonumber \\
 & & \times
 \sum\limits_{n=0}^\infty
 \frac{\Gamma(n+\nu_2)\Gamma(n-D/2+\nu_{123})}
      {\Gamma(n+1)\Gamma(n+\nu_{23})}
 \left(1-x\right)^n.
\eq
where $x=(p_1+p_2)^2/(p_1+p_2+p_3)^2$,
\bq
c_\Gamma & = &
\frac{\Gamma(1-2\varepsilon)}{\Gamma(1+\varepsilon) \Gamma(1-\varepsilon)^2}.
\eq
and the $\nu_j$ are the powers to which each propagator is raised.
For $\nu_1=\nu_2=\nu_3=1$ the expression simplifies and one obtains for
the Laurent expansion in $\eps$:
\bq
I(1,1,1) & = &
\frac{1}{\eps} \frac{\ln x}{(1-x)} - \frac{1}{2} \frac{\ln^2 x}{(1-x)}
+ {\cal O}(\eps)
\eq
From explicit higher order calculations it is emerging that 
one can express the results of Feynman integrals
in multiple polylogarithms.
Multiple polylogarithms are a generalization of the logarithm.
In particular they can depend on several scales $x_1$, $x_2$, ..., $x_k$,
as opposed to the logarithm, which only depends on one argument $x$.
Multiple polylogarithms have been studied in recent years by
mathematicians and physicists
\cite{Goncharov,Borwein,Remiddi:1999ew,Gehrmann:2000zt,Moch:2001zr}.
Multiple polylogarithms can either be defined by an integral representation
or a sum representation.
They satisfy two distinct Hopf algebras and it is convenient to introduce
both definitions to discuss the algebraic properties.
First the definition by an iterated integral representation:
\bq
G(z_1,...,z_k;y) & = & \int\limits_0^y \frac{dt_1}{t_1-z_1}
 \int\limits_0^{t_1} \frac{dt_2}{t_2-z_2} ...
 \int\limits_0^{t_{k-1}} \frac{dt_k}{t_k-z_k},
 \nonumber \\
G(0,...,0;y) & = & \frac{1}{k!} \left( \ln y \right)^k.
\eq
For fixed $y$ the functions $G(z_1,...,z_k;y)$ satisfy a shuffle algebra
in the letters $z_1$, ..., $z_k$.
An example for the multiplication is:
\bq
G(z_1;y) G(z_2;y) & = & G(z_1,z_2;y) + G(z_2,z_1;y) 
\eq
Alternatively multiple polylogarithms can be defined by
an iterated sum representation:
\bq
\mbox{Li}_{m_k,...,m_1}(x_k,...,x_1)
 & = &
\sum\limits_{i_1>i_2>\ldots>i_k>0}
     \frac{x_1^{i_1}}{{i_1}^{m_1}}\ldots \frac{x_k^{i_k}}{{i_k}^{m_k}}.
 \nonumber 
\eq
The functions $\mbox{Li}_{m_k,...,m_1}(x_k,...,x_1)$ satisfy 
a quasi-shuffle algebra
in the letters ${x_j}^i/i^{m_j}$.
An example for the multiplication is:
\bq
\lefteqn{
\mbox{Li}_{m_1}(x_1) \mbox{Li}_{m_2}(x_2) = }
 \nonumber \\
 & &  
 \mbox{Li}_{m_1,m_2}(x_1,x_2)
+\mbox{Li}_{m_2,m_1}(x_2,x_1)
+\mbox{Li}_{m_1+m_2}(x_1 x_2).
 \nonumber 
\eq
Note the additional third term on the r.h.s. as compared to a shuffle product.
A quasi-shuffle algebra is also a Hopf algebra \cite{Hoffman}.

The functions $G(z_1,...,z_k;y)$ and $\mbox{Li}_{m_k,...,m_1}(x_k,...,x_1)$
denote the same class of functions.
With the short-hand notation
\bq
\lefteqn{
G_{m_1,...,m_k}(z_1,...,z_k;y) = }
 \nonumber \\
 & &
 G(\underbrace{0,...,0}_{m_1-1},z_1,...,z_{k-1},\underbrace{0...,0}_{m_k-1},z_k;y)
 \nonumber 
\eq
the relation between the two notations is given by
\bq
\lefteqn{
\mbox{Li}_{m_k,...,m_1}(x_k,...,x_1) = }
 \nonumber \\
 & & 
 (-1)^k 
 G_{m_1,...,m_k}\left( \frac{1}{x_1}, \frac{1}{x_1 x_2}, ..., \frac{1}{x_1...x_k};1 \right).
\eq
The two notations are introduced to exhibit
the two different Hopf algebra structures.

Multiple polylogarithms were needed for the two-loop calculation
for $e^+ e^- \rightarrow 3\; \mbox{jets}$.
Here,
classical polylogarithms like $\mbox{Li}_{n}(x)$ are not
sufficient, since there are two variables inherent in the problem:
\bq
x_1 = \frac{s_{12}}{s_{123}}, 
 & &
x_2 = \frac{s_{23}}{s_{123}}.
\eq 
The calculation has been performed independently by two groups,
one group used shuffle algebra relations from the integral representation
\cite{Garland:2001tf,Garland:2002ak},
the other group used the
quasi-shuffle algebra from the sum representation
\cite{Moch:2002hm}.
Although these calculations exploited mainly properties related to the
multiplication in the two algebras, the co\-algebraic properties can be
used to simplify expressions \cite{Weinzierl:2003jx}.
Integration-by-part identities relate
the combination $G(z_1,...,z_k;y) + (-1)^k G(z_k,...,z_1;y)$
to $G$-functions of 
low\-er depth:
\bq
\label{ibp}
\lefteqn{
G(z_1,...,z_k;y) + (-1)^k G(z_k,...,z_1;y) } & & \nonumber \\
& = & G(z_1;y) G(z_2,...,z_k;y) - G(z_2,z_1;y) G(z_3,...,z_k;y)
 + ... \nonumber \\
& &
 - (-1)^{k-1} G(z_{k-1},...z_1;y) G(z_k;y),
\eq
Eq. (\ref{ibp}) can also be derived in a different way.
In a Hopf algebra we have for any non-trivial element $w$ 
the following relation involving the antipode:
\bq
\label{axiomantipode}
w^{(1)} \cdot {\cal S}( w^{(2)} ) & = & 0.
\eq
Working out the relation (\ref{axiomantipode}) for the shuffle 
algebra of the functions
$G(z_1,...,$ $z_k;y)$, we recover (\ref{ibp}).
A similar relation can be obtained for the functions
$\mbox{Li}_{m_k,...,m_1}(x_k,...,x_1)$ using the quasi-shuffle algebra.

\section{Summary}
\label{sect:summary}

Hopf algebras occur in physics within the domains of 
quantum groups, integrable systems and quantum field theory.
In this talk I focussed on the last point and discussed 
occurrences in perturbation theory.
Hopf algebras allow to express certain combinatorial properties in a clean way.
I discussed the reformulation of the forest formula for renormalization,
Wick's theorem and its relation to deformed products
as well as the equivalence of relations obtained from 
integration-by-parts and the antipode in the case of Feynman loop integrals.
The underlying Hopf algebras are all commutative, 
but not necessarily cocommutative.

\end{document}